\documentclass[aps,prl,twocolumn,superscriptaddress,showpacs]{revtex4}

\usepackage{graphicx}
\begin{document}


\title{ Strong magnetic field enhancement of spin triplet pairing 
arising from \\ coexisting $2k_F$ spin and $2k_F$ charge fluctuations}



\author{Hirohito Aizawa}
\affiliation{
 Department of Applied Physics and Chemistry, The University of
 Electro-Communications, Chofu, Tokyo 182-8585, Japan}

\author{Kazuhiko Kuroki}
\affiliation{
 Department of Applied Physics and Chemistry, The University of
 Electro-Communications, Chofu, Tokyo 182-8585, Japan}

\author{Yukio Tanaka}
\affiliation{
 Department of Applied Physics, Nagoya University,
 Nagoya, 464-8603, Japan}


\date{\today}

\begin{abstract}
We study the effect of the magnetic field (Zeeman splitting) 
on the triplet pairing. We show generally that the enhancement of 
spin triplet pairing mediated by coexisting $2k_F$ spin and $2k_F$ charge 
fluctuations can be much larger than in the case of triplet pairing 
mediated by ferromagnetic spin fluctuations.
We propose that this may be related to the recent experiment for 
(TMTSF)$_2$ClO$_4$, in which a possibility of singlet to triplet pairing 
transition has been suggested.
\end{abstract}

\pacs{74.70.Kn, 74.20.Rp, 74.20.Mn}

\maketitle

 Spin triplet superconductivity is one of the most fascinating
 unconventional superconducting state. 
 The investigation of the mechanism of such a pairing state has been 
 a theoretical challenge.
 Spin triplet pairing mediated by ferromagnetic spin fluctuations has been
 studied from the early days in the context of 
 superfluid $^{3}$He,
 but another possibility has arisen 
 for the past several years: triplet pairing mediated by 
 coexisting $2k_F$ spin and $2k_F$ charge (or orbital) fluctuations 
 proposed for Sr$_2$RuO$_4$\cite{Takimoto} and 
 for organic superconductors (TMTSF)$_2$X (X=PF$_{6}$, ClO$_{4}$, etc.)
\cite{KKrev, Kuroki-Arita-Aoki, Tanaka-Kuroki, Kuroki-Tanaka, 
 Fuseya-Suzumura, Nickel-Duprat} and $\theta$-(BEDT-TTF)$_2$I$_3$.
\cite{KKtheta} 

For (TMTSF)$_2$X in particular, 
the possibility of spin triplet pairing has been 
pointed out starting from the early days 
\cite{Abrikosov, Takigawa-Yasuoka,Hasegawa-Fukuyama, Lebed, Dupuis, 
 Lee-Naughton, Lee-Chaikin, Lee-Brown, Lee-Chow, 
 Oh-Naughton, Shinagawa-Wu},  and two of the present authors as well 
as several other groups have come up with the possibility of 
a close competition between singlet $d$-wave-like 
pairing and a triplet $f$-wave-like pairing due to coexisting 
$2k_F$ spin and $2k_F$ charge fluctuations.
Very recently,  an NMR study on (TMTSF)$_{2}$ClO$_{4}$ has pointed out 
 a possibility of a transition from a spin singlet pairing 
at low magnetic fields
 to triplet pairing or an FFLO state at high fields\cite{Shinagawa-Kurosaki}.
 In fact, such a possibility of singlet to triplet pairing 
 transition under high magnetic field was pointed out theoretically
\cite{Shimahara, Vaccarella-Melo, Fuseya-Onishi}.

 In the present Letter, 
we study the effect of the magnetic field (Zeeman splitting) 
on the triplet pairing using random phase approximation (RPA). 
We show generally that the enhancement of 
spin triplet pairing mediated by coexisting $2k_F$ spin and $2k_F$ charge 
fluctuations can be much larger than in the case of triplet pairing 
mediated by ferromagnetic spin fluctuations.
Applying the idea to a microscopic model  for (TMTSF)$_2$X  
in which strong $2k_F$ spin and $2k_F$ charge fluctuations take place, 
we actually show that 
 the magnetic field enhancement of spin triplet $f$-wave pairing 
is strong compared to 
 the enhancement of ferromagnetic-spin-fluctuation mediated 
 triplet pairing that occurs in a triangular lattice Hubbard model
\cite{Kuroki-Arita, Arita-Kuroki}.
 Due to this strong effect, 
we show that even when spin singlet pairing dominates in the 
 absence of the magnetic field, a transition to triplet pairing 
 may take place by applying the magnetic field.
 This is consistent with the possibility 
 of the magnetic field induced singlet-triplet transition in 
 (TMTSF)$_2$ClO$_4$\cite{Shinagawa-Kurosaki}. Moreover, 
this strong magnetic field effect may be used as a general 
probe for identifying the pairing mechanism of 
triplet superconductors.


 The extended Hubbard Hamiltonian that takes into account the Zeeman effect
is given as
 $H = \sum_{i,j,\sigma} t_{ij} c_{i \sigma}^\dagger c_{j \sigma}
   + \sum_{i} U n_{i \uparrow} n_{i \downarrow}
   + \sum_{i,j,\sigma,\sigma'} V_{ij} n_{i \sigma} n_{j \sigma'}
   +g\mu_{B}B \sum_{i,\sigma} {\rm sgn}(\sigma)
   c_{i \sigma}^\dagger c_{i \sigma}$,
%
%
 where $c_{i \sigma}^\dagger$ creates an electron with spin $\sigma$ at
the  $i$-th site.
 $t_{ij}$ represents the hopping parameters, $U$ is the on-site interaction
 and $V_{ij}$ are the off-site interactions.
 $g\mu_{B}B$ is the Zeeman energy with the spin quantization axis
 $\hat{z} \parallel {\bf B}$.
We ignore the orbital effect of the magnetic field.
 Within RPA, the effective pairing interactions 
mediated by spin and  charge fluctuations 
are given as 
 \begin{eqnarray}
  V^{s} \left({\bf q}\right) &=& U+V\left({\bf q}\right)
   +\frac{U^{2}}{2} \chi_{sp}^{zz}\left({\bf q}\right)
   +U^{2}\chi_{sp}^{+-}\left({\bf q}\right)
   \nonumber \\ & &
   -\frac{\left(U+2V\left({\bf q}\right)\right)^{2}}{2}
   \chi_{ch}\left({\bf q}\right),
 \label{Vs}
 \end{eqnarray}
 for spin singlet pairing,
 \begin{eqnarray}
  V^{t \sigma \sigma} \left({\bf q}\right) &=& V\left({\bf q}\right)
   -2\left(U+V\left({\bf q}\right)\right)V\left({\bf q}\right)
   \chi^{{\sigma} \bar{\sigma}}
   \nonumber \\ & &
   -V\left({\bf q}\right)^{2}\chi^{{\sigma} {\sigma}}
   -\left(U+V\left({\bf q}\right)\right)^{2}
   \chi^{\bar{\sigma} \bar{\sigma}},
 \label{Vt-para}
 \end{eqnarray}
 for spin triplet pairing with ${\bf d} \perp \hat{z}$
 which means total $S_{z}=+1 (-1)$ for $\sigma=\uparrow (\downarrow)$,
 \begin{eqnarray}
  V^{t \sigma \bar{\sigma}} \left({\bf q}\right) &=& V\left({\bf q}\right)
   +\frac{U^{2}}{2} \chi_{sp}^{zz}\left({\bf q}\right)
   -U^{2}\chi_{sp}^{+-}\left({\bf q}\right)
   \nonumber \\ & &
   -\frac{\left(U+2V\left({\bf q}\right)\right)^{2}}{2}
   \chi_{ch}\left({\bf q}\right),
 \label{Vt-oppo}
 \end{eqnarray}
 for spin triplet pairing with
 ${\bf d} \parallel \hat{z}$ ($S_{z}=0$), 
 where $V\left({\bf q}\right)$ is 
 the Fourier transform of the off-site interactions.
 The longitudinal spin susceptibility and the charge susceptibility
 are obtained by
 $\chi_{sp}^{zz}=\left(\chi^{\uparrow\uparrow}+\chi^{\downarrow\downarrow}
 -\chi^{\uparrow\downarrow}-\chi^{\downarrow\uparrow}\right)/2$
 and 
 $\chi_{ch}=\left(\chi^{\uparrow\uparrow}+\chi^{\downarrow\downarrow}
 +\chi^{\uparrow\downarrow}+\chi^{\downarrow\uparrow}\right)/2$.
 Here,
 \begin{eqnarray}
 \chi^{\sigma \sigma} =
  \frac{\left(1+\chi_{0}^{{\bar \sigma} {\bar \sigma}}
	 V_{\bf q}\right)\chi_{0}^{\sigma \sigma}}
  {\left(1+\chi_{0}^{\sigma \sigma}
    V_{\bf q}\right)
  \left(1+\chi_{0}^{{\bar \sigma} {\bar \sigma}}
   V_{\bf q}\right)
  -\left(U+V_{\bf q}\right)^{2}
  \chi_{0}^{\sigma \sigma}\chi_{0}^{{\bar \sigma} {\bar \sigma}}}
 \label{chi-para}
 \end{eqnarray}
 and
 \begin{eqnarray}
  \chi^{\sigma {\bar \sigma}} =
   \frac{-\chi_{0}^{\sigma \sigma}
   \left(U+V_{\bf q}\right)\chi_{0}^{{\bar \sigma} {\bar \sigma}}}
   {\left(1+\chi_{0}^{\sigma \sigma}
     V_{\bf q}\right)
   \left(1+\chi_{0}^{{\bar \sigma} {\bar \sigma}}
    V_{\bf q}\right)
   -\left(U+V_{\bf q}\right)^{2}
   \chi_{0}^{\sigma \sigma}\chi_{0}^{{\bar \sigma} {\bar \sigma}}},
 \label{chi-anti-para}
 \end{eqnarray}
 where $V_{\bf q}$ stands for $V\left({\bf q}\right)$.
 The longitudinal bare susceptibility is given as
 \begin{eqnarray}
  \chi_{0}^{\sigma \sigma}\left( {\bf q}\right) =
   \frac{-1}{N}\sum_{{\bf k}}
   \frac{f\left( \xi_{\sigma}\left({\bf k}+{\bf q}\right) \right)
   -f\left( \xi_{\sigma}\left({\bf k}\right) \right)}
   {\xi_{\sigma}\left({\bf k}+{\bf q}\right)
   -\xi_{\sigma}\left({\bf k}\right)},
 \label{chi0-para}
 \end{eqnarray}
 where $\xi_{\sigma}\left({\bf k}\right)$ is
 the single electron dispersion that considers the Zeeman effect
 measured from the chemical potential $\mu$, and
 $f\left( \xi_{\sigma}\left({\bf k}\right)\right)$ is the Fermi distribution
 function.
 The transverse spin susceptibility 
 is given as
 $\chi_{sp}^{+-}=\chi_{0}^{+-}/(1-U\chi_{0}^{+-})$,
 where the transverse bare susceptibility is obtained by
\begin{eqnarray}
 \chi_{0}^{+-}\left( {\bf q}\right) =
  \frac{-1}{N}\sum_{{\bf k}}
  \frac{f\left( \xi_{\sigma}\left({\bf k}+{\bf q}\right) \right)
  -f\left( \xi_{{\bar \sigma}}\left({\bf k}\right) \right)}
  {\xi_{\sigma}\left({\bf k}+{\bf q}\right)
  -\xi_{{\bar \sigma}}\left({\bf k}\right)}.
 \label{chi0-trans}
\end{eqnarray}

 To obtain the superconducting state,
 we solve the linearized BCS gap equation within weak-coupling theory,
 \begin{eqnarray}
  \lambda^{\mu} \phi^{\mu} \left({\bf k}\right)
   &=& -\sum_{\bf k'}V^{\mu}\left({\bf k}-{\bf k'}\right)
   \nonumber \\ &\times&
   \frac
   {1-f\left(\xi_{\sigma}\left({\bf k'}\right)\right)
   -f\left(\xi_{\sigma'}\left({\bf k'}\right)\right)}
   {\xi_{\sigma}\left({\bf k'}\right)+\xi_{\sigma'}\left({\bf k'}\right)}
   \phi^{\mu} \left({\bf k'}\right).
   \label{gap-eq}
 \end{eqnarray}
We consider singlet and triplet pairings 
with ${\bf d} \parallel \hat{z}$ 
 ($\mu=s, t\sigma {\bar \sigma}$) for opposite spin pairing 
($\sigma\ne\sigma'$), 
 and triplet pairing with ${\bf d} \perp \hat{z}$ 
 ($\mu=t \sigma \sigma$)
 for parallel spin pairing($\sigma=\uparrow$, $\downarrow$).
 $\phi^{\mu} \left({\bf k}\right)$ is the gap function and
 the critical temperature $T_{c}$ is determined as the temperature
 where the eigenvalue $\lambda$ reaches unity.
 To give a reference for the values of the 
magnetic field, we calculate the Pauli limit
 by $\mu_{B}B_{P}=1.75k_{B}T_{c}/\sqrt{2}$.
 Although RPA may be considered as 
quantitatively insufficient for discussing the
 absolute value of $T_{c}$, we expect this approach to be valid for
 studying the {\it competition} between different pairing symmetries.
 In fact, as we shall see, we find very good agreement between the 
RPA results and the already known results obtained by 
dynamical cluster approximation (DCA)\cite{Arita-Kuroki}.

Before giving the calculation results, we show generally 
using the above formula that the effect of the Zeeman splitting 
on the triplet pairing caused by the coexistence of $2k_F$ spin and 
$2k_F$ charge fluctuations can be very special.
First, let us consider a case where off-site repulsions are not present,
so that only the spin fluctuations are relevant, and therefore 
possibility of triplet pairing superconductivity arises due to 
ferromagnetic spin fluctuations.
In this case, the triplet pairing 
interactions reduce to 
 $V^{t \sigma \sigma} \left({\bf q}\right)=
-U^{2} \chi^{\bar{\sigma} \bar{\sigma}}$ and 
 $V^{t \sigma \bar{\sigma}} \left({\bf q}\right)=
+\frac{U^{2}}{2} \chi_{sp}^{zz}\left({\bf q}\right)
-U^{2}\chi_{sp}^{+-}\left({\bf q}\right)$,
where the formula for $\chi^{\sigma \sigma}$ also reduces to  
\begin{equation}
\chi^{\sigma\sigma}=\frac{\chi_{0}^{\sigma\sigma}}
{1-U^2\chi_{0}^{\sigma\sigma}\chi_{0}^{{\bar \sigma}{\bar \sigma}}}.
\label{chi-dd-red}
\end{equation}
Here, we assume without losing generality that 
$\chi_{0}^{\sigma\sigma}$ is enhanced while 
$\chi_{0}^{{\bar \sigma}{\bar \sigma}}$ is suppressed by the 
magnetic field. (Whether $\sigma=\uparrow$ or $\sigma=\downarrow$ 
depends on the band structure and the band filling of the system
as we shall see later.)
In the first order of the magnetic field,   
$\chi_{sp}^{zz}$ and $\chi_{sp}^{+-}$ are not affected because 
exchanging $\uparrow$ and $\downarrow$ do not affect these quantities.
$\chi^{\sigma\sigma}$ is enhanced because the 
numerator $\chi_{0}^{\sigma\sigma}$  
in eq.(\ref{chi-dd-red}) is enhanced, but again 
the term in the denominator, 
$\chi_{0}^{\sigma \sigma}\chi_{0}^{\bar{\sigma}\bar{\sigma}}$, 
is not affected by the magnetic field in its first order.
Thus, although $V^{t\bar{\sigma} \bar{\sigma}}$ should dominate over 
$V^{t\bar{\sigma} \sigma}$, its enhancement due to the Zeeman splitting 
occurs only through the direct enhancement of 
$\chi_{0}^{\sigma\sigma}$, which may not be so large 
for realistic magnetic fields.

\begin{figure}[!htb]
  \includegraphics[width=8cm]{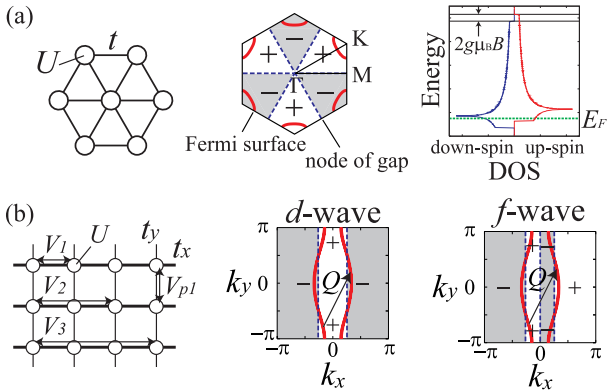}
  \caption{(color online)
  (a) The model on the triangular lattice (left), 
  the $f$-wave gap function (center), and the density of
  states in the presence of the Zeeman splitting schematically, 
 the Fermi level is for a dilute band filling (right). 
  (b) The model for (TMTSF)$_2$X on a Q1D lattice (left), and 
  $d$-wave (center) and $f$-wave (right) gap functions 
 along with the Fermi surface The arrows represent 
 the nesting vector of the Fermi surface.}
  \label{model-gap}
 \end{figure}

When a possibility of triplet pairing arises due to 
the coexistence of $2k_F$ spin and $2k_F$ charge fluctuations, 
where the latter are induced by off-site repulsions, 
the situation can change drastically.
To make the discussion simple, let us consider a case with 
$-(U+2V({\bf Q}_{2k_F}))\simeq U$, or equivalently 
$U+V({\bf Q}_{2k_F})\simeq 0$, 
for which, in the absence of magnetic field,  
$\chi_{sp}^{zz}({\bf Q}_{2k_F})=\chi_{sp}^{+-}({\bf Q}_{2k_F})\simeq 
\chi_{ch}({\bf Q}_{2k_F})$ and thus 
$V^{t\sigma\sigma}({\bf Q}_{2k_F})=V^{t\sigma\bar{\sigma}}({\bf Q}_{2k_F})
\simeq -V^s({\bf Q}_{2k_F})$.
(We shall see later that our idea works for more general cases).
Here, the singlet and the triplet pairing interactions have nearly the same 
absolute values because the 
$2k_F$ spin and the $2k_F$ charge contributions  
work constructively (destructively) in the spin triplet (singlet) pairing 
interaction.\cite{Takimoto, Kuroki-Arita-Aoki, Tanaka-Kuroki, Kuroki-Tanaka, 
Fuseya-Suzumura,Nickel-Duprat}.
In this case, $\chi^{\sigma\sigma}$ at ${\bf q=Q}_{2k_F}$
can be given by a reduced form 
\begin{equation}
\chi^{\sigma\sigma}({\bf Q}_{2k_F})\simeq
\frac{\chi_0^{\sigma\sigma}({\bf Q}_{2k_F})}
{1+V({\bf Q}_{2k_{F}}) \chi_0^{\sigma\sigma}({\bf Q}_{2k_F})}.
\label{chi-v-reduced}
\end{equation}
Here again, we assume without losing generality that 
$\chi_0^{\sigma\sigma}({\bf Q}_{2k_F})$
($\chi_0^{\bar{\sigma} \bar{\sigma}}({\bf Q}_{2k_F})$) is 
enhanced (suppressed) by the Zeeman splitting.
$\chi_{sp}^{zz}$ and $\chi_{sp}^{+-}$ again are not 
affected by the magnetic field in its first order, 
so $V^{s}$ and $V^{t\sigma\bar{\sigma}}$ 
are unaffected, while $\chi^{\sigma\sigma}$ is again affected.
The difference from the case with ferromagnetic spin fluctuations 
lies in that 
since $V({\bf Q}_{2k_F})<0$, the denominator of eq.(\ref{chi-v-reduced}) 
decreases as $\chi_0^{\sigma\sigma}$
increases. The enhancement of $\chi^{\sigma\sigma}$ due to this 
effect can be very large in the 
vicinity of $2k_F$ CDW ordering because 
$1+(U+2V({\bf Q}_{2k_F}))\chi_0({\bf Q}_{2k_F})=0$ signals this ordering,
which is the same as 
$1+V({\bf Q}_{2k_F})\chi_0({\bf Q}_{2k_F})=0$ when 
$U+V({\bf Q}_{2k_F})=0$.
Therefore, when the possibility of triplet pairing arises in the 
vicinity of coexisting $2k_F$ CDW and $2k_F$ SDW phases, 
triplet pairing with parallel spins can be strongly favored by the 
Zeeman splitting. In actual cases, superconductivity is usually 
degraded by the orbital effect under magnetic fields, 
but even in that case, the enhancement of 
triplet pairing due to the above effect should make the suppression to 
be moderate. For quasi-1D systems in particular, where the additional 
node in the SC gap required in the triplet pairing does not intersect 
the Fermi surface (see Fig.\ref{model-gap}(b)), the coexistence of 
$2k_F$ spin and $2k_F$ charge fluctuations already results in a subtle 
competition between singlet and triplet pairings 
\cite{ Tanaka-Kuroki, Kuroki-Tanaka}, 
so that the strong enhancement of the triplet pairing interaction 
by the magnetic field may easily result in 
a singlet to triplet pairing transition.

We now apply the above idea to actual systems.
First, we consider a case where the possibility of 
triplet pairing occurs due to 
ferromagnetic spin fluctuations induced by the on-site repulsion.
As a typical example, we consider a case on a triangular lattice 
with dilute band filling as shown in Fig. \ref{model-gap}(a).
In this case, possibility of spin triplet $f$-wave pairing 
due to ferromagnetic spin fluctuations 
has been pointed out previously\cite{ Kuroki-Arita, Arita-Kuroki}. 

\begin{figure}[!htb]
 \centering
 \includegraphics[width=8cm]{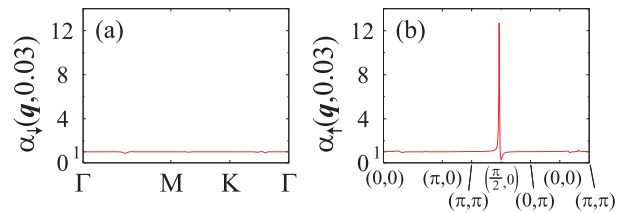}
 \caption{(color online)
 $\alpha_\sigma({\bf q},B)$ at $T=0.02$ for 
 (a)the triangular lattice with $n=0.2$ and 
 (b)the Q1D model with $V_{2}+V_{p1}=0.85=U/2$.}
 \label{ratio-chi0}
\end{figure}

\begin{figure}[!htb]
 \centering
 \includegraphics[width=8cm]{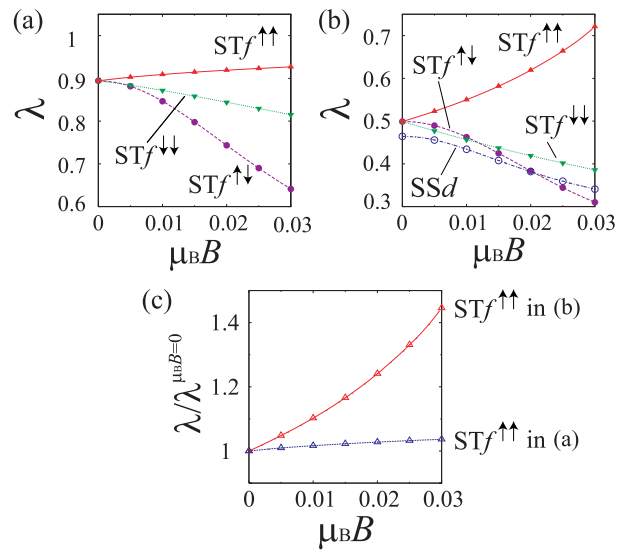}
 \caption{(color online)
 The $B$ dependence of $\lambda$ at $T=0.02$ for 
 (a)the triangular lattice for $n=0.2$, 
 (b)the Q1D model for $V_{2}+V_{p1}=U/2$.
 (c) The $B$ dependence of the most dominant $\lambda$ for the 
cases shown in (a) and (b) normalized by $\lambda$ at $B=0$.
}
 \label{muBB-lambda}
\end{figure}

 The band dispersion is given as 
 $\xi_{\sigma}\left({\bf k}\right)
 =2t\cos k_{x}+2t\cos k_{y}+2t\cos(k_{x}+k_{y})
 -\mu+g\mu_{B}B\rm{sgn}(\sigma)$.
 We take the transfer energy as the unit of the energy, i.e., $t=1.0$
 The on-site interaction is $U=3.0$, and 
the band filling is taken as $n=0.2$.
 We take  $128 \times 128$ $k$-point meshes in the RPA calculation.
 When the magnetic field is absent,
 we obtain $k_{B}T_{c} \simeq 0.014$.
 The Pauli limit corresponding to this $T_c$ is $\mu_{B}B_{P} \simeq 0.017$,
which should be considered as a reference for the values of the 
magnetic field.
 When the Zeeman splitting is introduced,
 $\chi_0^{\downarrow \downarrow}$ becomes slightly larger than
 $\chi_0^{\uparrow \uparrow}$ because of the 
the increase of the density of states (DOS) at the Fermi level of the 
down-spin  due to the Zeeman splitting (see Fig.\ref{model-gap}(a)).
Thus, this corresponds to the case with $\sigma=\downarrow$ in our 
general argument for the case with ferromagnetic fluctuations, 
so that we should focus on the enhancement of 
$\chi^{\downarrow\downarrow}$.
In order to measure how the 
magnetic field enhancement of $\chi_0^{\downarrow \downarrow}$ 
is reflected to the enhancement of $\chi^{\downarrow\downarrow}$, we 
introduce a parameter  
$\alpha_\sigma({\bf q},B)=\frac
{\chi^{\sigma\sigma}({\bf q},B)/\chi^{\sigma\sigma}({\bf q},0)}
{\chi_{0}^{\sigma\sigma}({\bf q},B)/\chi_{0}^{\sigma\sigma}({\bf q},0)}$. 
In Fig.\ref{ratio-chi0}, 
$\alpha_\downarrow$ at $\mu_B B=0.03$ 
is plotted as a function 
of ${\bf q}$. $\alpha_\downarrow\simeq 1$ means that the effect of 
the denominator in eq.(\ref{chi-dd-red}) is small, as expected from our 
argument above. 
Effect of the magnetic field on the strength of the 
triplet pairing is shown in Fig.\ref{muBB-lambda}(a), where we plot the 
eigenvalues of the linearized gap equation for each triplet pairings.
As expected, due to the enhancement in $\chi^{\downarrow\downarrow}$, 
triplet $f^{\uparrow\uparrow}$-wave dominates in the presence of the magnetic 
field. We note that this result for $\lambda$ 
closely resembles the field dependence of the pairing susceptibility 
calculated by DCA for the 
same system\cite{Arita-Kuroki}, suggesting the reliability of the 
present approach.

We now turn to the case where the possibility of 
triplet pairing arises due to the coexistence of $2k_F$ spin 
and $2k_F$ charge fluctuations. As a typical example, we consider the 
case of (TMTSF)$_2$X.
\cite{ Kuroki-Arita-Aoki, Tanaka-Kuroki, Kuroki-Tanaka, 
Fuseya-Suzumura, Nickel-Duprat}
 We adopt a 3/4-filled Q1D extended Hubbard model as shown in 
Fig.\ref{model-gap}(b).\cite{Kuroki-Tanaka}
The band dispersion is
 given by $\xi_{\sigma}\left({\bf k}\right)=2t_{x}\cos k_{x}+2t_{y}\cos k_{y}
 -\mu+g\mu_{B}B\rm{sgn}(\sigma)$.
 We take $t_{x}=1.0$ as the unit of the energy, 
and $t_{y}=0.2$. 
 As for the interaction parameters, we consider not only
 the on-site interaction $U$ but also off-site interactions : 
 nearest-neighbor(n.n.) $V_{1}$, 2nd n.n. $V_{2}$,
 3rd n.n. interaction $V_{3}$ in the $x$-direction
 and n.n. interaction $V_{p1}$ in the $y$-direction, 
 where the Fourier transformed off-site interaction is given as
 $V\left({\bf q}\right)=2V_{1}\cos\left(k_{x}\right)
 +2V_{2}\cos\left(2k_{x}\right)+2V_{3}\cos\left(3k_{x}\right)
 +2V_{p1}\cos\left(k_{y}\right)$.
 We set the on-site and the off-site interactions as 
 $U=1.7$, $V_{1}=0.9$ and $V_{3}=0.1$, and 
 $V_{2}$ and $V_{p1}$ are varied.
 The band filling is taken as $n=1.5$ in accord with (TMTSF)$_{2}$X.
 We take  $256 \times 128$ k-point meshes in the actual calculation.
 We first consider the case with $V_{2}=0.45$, $V_{p1}=0.4$,
 for which $U+V({\bf Q}_{2k_F})=0$ (${\bf Q}_{2k_F}\simeq(\pi/2, \pi)$)
 and thus singlet $d$-wave and 
 triplet $f$-wave pairings are nearly degenerate in the absence of the 
 magnetic field. As discussed in the previous studies
\cite{Tanaka-Kuroki,Kuroki-Tanaka,Fuseya-Suzumura,Nickel-Duprat}, 
the degeneracy is because 
the triplet and singlet pairing interactions are nearly  
equal at ${\bf Q}_{2k_F}$, and 
the additional gap node in the $f$-wave pairing does not intersect the 
Fermi surface, so that the nodal structure on the Fermi surface is the 
same between $d$ and $f$, as shown in Fig.\ref{model-gap}(b).
Introduction of the Zeeman splitting 
enhances $\chi_0^{\uparrow\uparrow}({\bf Q}_{2k_F})$ and suppresses 
$\chi_0^{\downarrow\downarrow}({\bf Q}_{2k_F})$  
because the up-spin band 
becomes close to half filling and the electron-hole symmetry is more restored.
Thus, this corresponds to 
$\sigma=\uparrow$ in our general argument for the case with 
coexisting $2k_F$ spin and $2k_F$ charge fluctuations, so 
we should look at the enhancement of $\chi^{\uparrow\uparrow}$ 
due to the magnetic field.
In Fig.\ref{ratio-chi0}(b), 
$\alpha_\uparrow$ is plotted at $\mu_BB=0.03$, which 
largely exceeds unity at ${\bf Q}_{2k_F}$ as expected from our 
previous argument that the denominator in eq.(\ref{chi-v-reduced}) becomes 
close to 0. In Fig.\ref{muBB-lambda}(b), 
we show the magnetic field dependence of the 
eigenvalue $\lambda$. It can be seen that triplet 
$f^{\uparrow\uparrow}$-wave 
is strongly enhanced due to the strong enhancement of 
$\chi^{\uparrow\uparrow}({\bf Q}_{2k_F})$. 
In Fig.\ref{muBB-lambda}(c), 
we compare the $B$ dependence of 
$\lambda$ normalized by its value at $B=0$ for 
the two cases. We can see that the enhancement of the 
triplet pairing mediated by coexisting $2k_F$ spin and $2k_F$ charge 
fluctuations is much larger.

\begin{figure}[!htb]
 \centering
 \includegraphics[width=8cm]{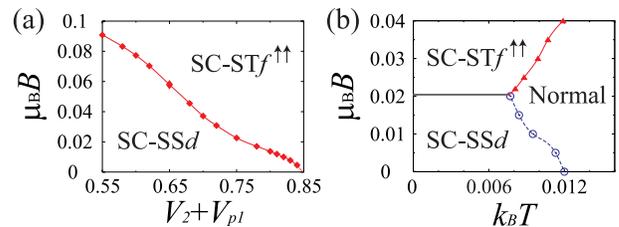}
 \caption{(color online)
 (a) Pairing phase diagram at $k_{B}T=0.012$.
 (b) Calculated $T$-$B$ phase diagram in $V_{2}=0.4$ and $V_{p1}=0.4$.}
 \label{V2Vp1-muB-phase}
\end{figure}

In the above, we considered a case where 
$\chi_{ch}({\bf Q}_{2k_F})=\chi_{sp}({\bf Q}_{2k_F})$ and thus 
the singlet $d$-wave and triplet $f$-wave pairings 
are nearly degenerate at $B=0$, but 
even when singlet pairing dominates at $B=0$
a transition to triplet pairing 
can take place within realistic values of $B$ 
due to this strong enhancement of the triplet pairing interaction. 
To see this in detail, we consider a case with 
$V_{2}=0.4$, $V_{p1}=0.4$, for which $d$-wave 
dominates over $f$-wave for $B=0$.
Here, $T_c=0.012$, which    
corresponds to the Pauli limit of $\mu_{B}B_{P} \simeq 0.015$.
 We obtain in Fig. \ref{V2Vp1-muB-phase}(a) 
a pairing ``phase diagram'' in the $(V_2+V_{p1})$-$B$ 
 space obtained by comparing $\lambda$ at $k_{B}T=0.012$.
 The phase diagram for the superconducting
 state(spin singlet $d$-wave and triplet $f^{\uparrow \uparrow}$-wave)
 and the normal state in the temperature-magnetic field space
 is shown in Fig. \ref{V2Vp1-muB-phase}(b).
 Applying the magnetic field,
 the spin singlet $d$-wave (indicated by SC-SS$d$) gives way to
 the spin triplet $f$-wave pairing with $S_{z}=+1$
 (SC-ST$f^{\uparrow \uparrow}$).
 Note that the $T_{c}$ in the spin triplet
 $f^{\uparrow \uparrow}$-wave channel increases with $B$
because we ignore the orbital effect.
 We expect that  the orbital effect actually suppresses the 
$T_c$, but even in that case, the effect of the Zeeman splitting should 
strongly favor the occurrence of triplet pairing over singlet pairing.

 In conclusion, we have generally shown that 
 the magnetic field enhancement of the spin triplet pairing 
 due to the coexistence of $2k_F$ spin and $2k_F$ charge fluctuations 
 can be extremely large compared to that mediated by ferromagnetic spin 
 fluctuations. Thus, even when spin singlet pairing dominates in the 
 absence of the magnetic field, a transition to spin triplet pairing 
 can take place by applying (not unrealistically large) magnetic field.
 This is consistent with the possibility 
 of the magnetic field induced singlet-triplet transition in 
 (TMTSF)$_2$ClO$_4$\cite{Shinagawa-Kurosaki}.
\begin{acknowledgments}
 We acknowledge Grants-in-Aid for Scientific Research from the
 Ministry of Education, Culture, Sports, Science and Technology of
 Japan, and from the Japan Society for the Promotion of Science.
 Part of the calculation has been performed at the 
 facilities of the Supercomputer Center, 
 ISSP, University of Tokyo.
\end{acknowledgments}


\end{document}